# Analytic derivation of electrostrictive tensors and their application to optical force density calculations

Wujiong Sun[1,2], S. B. Wang[2], Jack Ng[3], Lei Zhou[1], and C. T. Chan[2*]

[1] State Key Laboratory of Surface Physics, Key Laboratory of Micro and Nano Photonic Structures (Ministry of Education), and Collaborative Innovation Center of Advanced Microstructures, Fudan University, Shanghai 200433, China
[2] Department of Physics and Institute for Advanced Study, The Hong Kong University of Science and Technology, Clear Water Bay, Hong Kong
[3] Department of Physics and Institute of Computational and Theoretical Studies, Hong Kong Baptist University, Hong Kong

## ABSTRACT

Using multiple scattering theory, we derived for the first time analytical formulas for electrostrictive tensors for two dimensional metamaterial systems. The electrostrictive tensor terms are found to depend explicitly on the symmetry of the underlying lattice of the metamaterial and they also depend explicitly on the direction of a local effective wave vector. These analytical results enable us to calculate light induced body forces inside a composite system (metamaterial) using the Helmholtz stress tensor within the effective medium formalism in the sense that the fields used in the stress tensor are those obtained by solving the macroscopic Maxwell equation with the microstructure of the metamaterial replaced by an effective medium. Our results point to some fundamental questions of using an effective medium theory to determine optical force density. In particular, the fact that Helmholtz tensor carries electrostrictive terms that are explicitly symmetry dependent means that the standard effective medium parameters cannot give sufficient information to determine body force density, even though they can give the correct total force. A more challenging issue is that the electrostrictive terms are related to a local effective wave vector, and it is not always obtainable in systems with boundary reflections within the context of a standard effective medium approach.

## I. INTRODUCTION

The optical properties of man-made materials consisting of an array of sub-wavelength elements can be described by effective medium parameters ($\varepsilon_{eff}$ and $\mu_{eff}$) if the operating wavelength is much bigger than the lattice constant of the microstructure. In the area of metamaterial research [1-6], it is frequently assumed that effective medium parameters provide sufficient information to determine the light-matter interaction. The novel light manipulation functionalities of these new materials, such as negative refractive indices and super-resolution imaging, are always explored and described using effective constitutive parameters. It is sometimes debatable whether effective medium parameters are really accurate for some experimentally fabricated samples because the lattice constant in real samples is not that small compared with the operation wavelength but in the limit that $\omega \to 0$ and $k \to 0$ (the wavelength is very long compared with the lattice constant), the effective medium parameters should provide the information needed to determine the wave-manipulating properties. But even if the effective medium parameters can describe faithfully how the material can manipulate wave in the true long wavelength limit, can the same set of parameters describe faithfully how the wave can manipulate the material? For example, can $\varepsilon_{eff}$ and $\mu_{eff}$ determine the light-induced force and light-induced stress acting on a metamaterial consisting of an array of sub-wavelength elements? For the sake of convenience, we will use the generic term "metamaterial" throughout this paper although the prototypical system we consider is actually a two-dimensional (2D) photonic crystal with a lattice constant that is very small compared with the wavelength so that the effective medium parameters can be made to be as accurate as we want as far as light scattering is concerned.

It is quite well known that the total electromagnetic induced force acting on a piece of metamaterial illuminated by an external light source can be calculated using the Maxwell stress tensor [7] if the light scattering property of the metamaterial can be described by standard effective medium parameters. Once the incident field is specified, we just need to solve a scattering problem to determine the scattering field and then an integration of the standard Maxwell stress tensor over a boundary enclosing the object will give the total optical force acting on the object. The procedure is conceptually straightforward although the computation can be challenging. However, it is considerably more difficult if we want to know the distribution of the light induced force density inside the metamaterial. We will see that we cannot use the Maxwell stress tensor to compute the force density and the standard effective medium parameters do not provide sufficient information to determine the force density even though they can provide enough information to solve the scattering problem.

In order to calculate the electromagnetic field induced force density inside an object, we need to consider the momentum conservation in a closed volume $V$ bounded by a surface $S$. The momentum conservation law can be written as [7]

$$-\int_V \vec{f}\, \mathrm{d}\tau + \oint_S \mathrm{d}\vec{S} \cdot \vec{\vec{T}} = \frac{\mathrm{d}}{\mathrm{d}\tau} \int_V \vec{g}\, \mathrm{d}\tau \quad . \tag{1a}$$

The first term contains the body force density $\vec{f}$ we want to calculate and the integral over the volume gives the total force. The second term is an integration of the electromagnetic stress tensor over the boundary $S$ enclosing the volume $V$. The third term represents the increment rate of electromagnetic momentum in the volume. The Abraham and Minkowski controversy about the definition of electromagnetic momentum inside of the dielectric media has lasted more than one century [8-14]. The correct interpretation of $\vec{g}$ in Eq.(1a) and the associated expression of the stress tensor are related to the long standing controversy. In this paper, we consider the time-averaged force density in stationary bodies, with time harmonic incident fields and steady state configurations. Under these conditions, the third term in Eq.(1a) can be set to be zero. We thus have

$$\int_V \vec{f}\, \mathrm{d}\tau = \oint_S \mathrm{d}\vec{S} \cdot \vec{\vec{T}} \tag{1b}$$

If we transform the surface integral of the right hand side of Eq. (1b) into a volume integral, we can get formally $f_i = \partial T_{ik}/\partial x_k$, which seems to tell us that the force density is given directly by the stress tensor. However, there are multiple ways to express stress tensors $\vec{\vec{T}}$ so that $\partial T_{ik}/\partial x_k$ does not give the same $f_i$ at points inside the object while these different stress tensors give exactly the same total force when integrated over an external boundary. In particular, we will show below that the standard Maxwell tensor does not give correct body forces, although it gives the correct total force. Other formulations of stress tensors such as the Minkowski [15,16], Abraham [17,18] and Einstein-Laub [19] also cannot give correct body forces in metamaterials due to the omission of electrostriction terms [20,21] as will be discussed in later sections. We note that many authors have recently studied electromagnetic force densities [22-28] and related topics, but the importance or the relevance of electrostriction have been largely ignored.

## II. DERIVATION OF ELECTROSTICTIVE TENSORS

### A. Maxwell and Helmholtz tensors

The computation of optical force and stress requires the knowledge of electromagnetic stress tensors, which will be specified in the following. For time harmonic fields, the time-averaged Maxwell stress tensor is [7,15-16].

$$T_{\text{Maxwell,ik}} = \frac{1}{2}\text{Re}\left[\varepsilon_0 E_i E_k^* - \frac{1}{2}\varepsilon_0 E^2 \delta_{ik} + \mu_0 H_i H_k^* - \frac{1}{2}\mu_0 H^2 \delta_{ik}\right] \qquad , \qquad (2)$$

where $\delta_{ik}$ is the Kronecher delta function, $\varepsilon_0$, $\mu_0$ are the permittivity and permeability of vacuum, and Re[...] means the real part of [...]. Based on the principle of virtual work, Helmholtz formulated a stress tensor in the static limit [29]. The result can be extended for time-harmonic electromagnetic fields, the stress tensor for the case of isotropic dielectric solid can be generalized as [20, 30-37].

$$T_{\text{Helmholtz,ik}} = \frac{1}{2}\text{Re}\left[\varepsilon_0 \varepsilon_r E_i E_k^* - \frac{1}{2}\varepsilon_0\left(\varepsilon_r \delta_{ik} + \frac{\partial \varepsilon_r}{\partial u_{ik}}\right)E^2 + \mu_0 H_i H_k^* - \frac{1}{2}\mu_0 H^2 \delta_{ik}\right] \qquad , \qquad (3)$$

where $\varepsilon_r$ is the relative permittivity of media, and $u_{ik}$ is the strain tensor $u_{ik} = \left(\partial u_i/\partial x_k + \partial u_k/\partial x_i\right)/2$, with $\vec{u}(\vec{x})$ being a displacement vector [32]. The electrostrictive tensor $\partial \varepsilon_r/\partial u_{ik}$ [30-37] describes how stretching (diagonal components in $u_{ik}$) and shearing (off-diagonal components in $u_{ik}$) will change the permitivity of a solid [31]. For isotropic amorphous material, $\partial \varepsilon_r/\partial u_{ik}$ reduces to $-\rho\left(\partial \varepsilon_r/\partial \rho\right)$ [31], where $\rho$ is the mass density of the material. Here, we assume for sake of mathematical simplicity that the relative permeability $\mu_r = 1$ but Eq. (3) can be extended straightforwardly to materials with $\mu_r \neq 1$ by adding magnetostrictive terms.

We note that the Helmholtz tensor is the same as the Maxwell tensor in vacuum but they are not the same inside a material. We will show that the Helmholtz tensor is the correct stress tensor to use to obtain the body forces induced by electromagnetic waves. The Helmholtz tensor contains explicitly the $\partial \varepsilon_r/\partial u_{ik}$ terms which are not given in standard effective medium theories (EMT) [38, 39].

## B. The prototypical system configuration

For simplicity, we consider a 2D system which is a circular domain as shown in Fig. 1(a) and Fig. 1(b). This electrically large circular domain is made up of a composite material which has an underlying micro-structure in the form of a 2D photonic crystal in which each unit cell contains a dielectric cylinder (non-dispersive, non-dissipative, positive relative permittivity $\varepsilon_c = 8$, and $\mu_c = 1$) that is aligned along the z direction. We consider two types of lattice structure having different symmetries, namely the square (Fig. 1(a)) and the hexagonal lattice (Fig. 1(b)). The insets of Fig. 1(a) and Fig.1(b) are zoom-ins to expose the microscopic lattice structure of the underlying photonic crystal. We consider the cases in which the lattice constant of the photonic crystal is very small compared with the wavelength so that the scattering properties can be described by an effective permittivity $\varepsilon_{\text{eff}}$. And for this reason, the optical properties of the circular domain can be well described by a homogenous dielectric cylinder with a dielectric constant of $\varepsilon_{\text{eff}}$ as shown schematically in Fig. 1(c). We shall purposely choose the lattice constant of the photonic crystals so that both square and hexagonal arrangements give the same $\varepsilon_{\text{eff}}$ according to standard EMT.

### C. Analytic derivation of electrostrictive terms using multiple scattering theory

In the following, we will assume that the working wavelength is much larger than the lattice constant of the underlying photonic crystal so that the effective medium parameters, as derived from standard EMT, provide a very accurate description of wave scattering properties. In the long-wavelength limit, the polarization-dependent effective parameters of the system shown in Fig. 1 can be derived conveniently using various methods such as multiple scattering theory (MST) [39-40]. It is well known that for a small filling ratio $p$, the effective permittivity $\varepsilon_{\text{eff}} = 1 + p(\varepsilon_c - 1)$ for the E$_z$ polarization (E field along z-axis), and $\varepsilon_{\text{eff}} = -1 + 2/(pM - 1)$ for the H$_z$ polarization (H field along z-axis) where $M = (\varepsilon_c - 1)/(\varepsilon_c + 1)$. These expressions are accurate for small values of $p$. Corrections for high filling ratios can also be derived systematically using MST [40].

The Helmholtz tensor, as shown in Eq. (3), carries electrostrictive terms. We found that the electrostrictive terms can also be derived using the MST formalism rigorously. The procedures are tedious and are given in the Appendix. Here we focus on the results and their implications. For the E$_z$ polarization, the result is:

$$\frac{\partial \varepsilon_{\text{eff}}}{\partial u_{xx}} = \frac{\partial \varepsilon_{\text{eff}}}{\partial u_{yy}} = -\left(\varepsilon_{\text{eff}} - 1\right), \qquad \frac{\partial \varepsilon_{\text{eff}}}{\partial u_{xy}} = \frac{\partial \varepsilon_{\text{eff}}}{\partial u_{yx}} = 0 \qquad . \qquad (4)$$

We note that the electrostrictive terms for the E$_z$ polarization do not depend on the symmetry of the underlying lattice and they are completely specified by the usual effective parameter $\varepsilon_{\text{eff}}$.

On the other hand, the electrostrictive terms for the H$_z$ polarization are more complicated. In particular, the tensor components depend not only on the symmetry of the underlying lattice structure, but also depend on the direction of a local effective wave vector even in the $\omega \to 0$ and $k \rightarrow 0$ limit. In addition, it is also possible to derive high order corrections using scattering theory for higher filling ratio ($p$). After some calculations (see Appendix A), the electrostrictive tensor components for a photonic crystal with a square lattice are found to be

$$\frac{\partial \varepsilon_{\text{eff}}}{\partial u_{xx}}, \frac{\partial \varepsilon_{\text{eff}}}{\partial u_{yy}} = -\frac{\varepsilon_{\text{eff}}^2 - 1}{2} \pm 1.297 \frac{\left(\varepsilon_{\text{eff}} - 1\right)^2}{2} \cos 2\phi_{K_{\text{eff}}} \, , \tag{5a}$$

$$\frac{\partial \varepsilon_{\text{eff}}}{\partial u_{xy}} = \frac{\partial \varepsilon_{\text{eff}}}{\partial u_{yx}} = -0.596 \frac{\left(\varepsilon_{\text{eff}} - 1\right)^2}{2} \sin 2\phi_{K_{\text{eff}}} \qquad . \tag{5b}$$

While that for a hexagonal lattice they have the form

$$\frac{\partial \varepsilon_{\text{eff}}}{\partial u_{xx}}, \frac{\partial \varepsilon_{\text{eff}}}{\partial u_{yy}} = -\frac{\varepsilon_{\text{eff}}^2 - 1}{2} \pm 0.5 \frac{\left(\varepsilon_{\text{eff}} - 1\right)^2}{2} \cos 2\phi_{K_{\text{eff}}} \qquad , \tag{6a}$$

$$\frac{\partial \varepsilon_{\text{eff}}}{\partial u_{xy}} = \frac{\partial \varepsilon_{\text{eff}}}{\partial u_{yx}} = \frac{\left(\varepsilon_{\text{eff}} - 1\right)^2}{2} \sin 2\phi_{K_{\text{eff}}} \qquad . \tag{6b}$$

In Eqs.(5a) and (6a), the "+" and "-" sign are for the $\partial \varepsilon_{\text{eff}} / \partial u_{xx}$ and $\partial \varepsilon_{\text{eff}} / \partial u_{yy}$ components, respectively. Corrections for high filling ratio can be derived systematically, and they are only required for diagonal terms. For the square lattice,

$$\frac{\partial \varepsilon_{\text{eff}}}{\partial u_{xx}}, \frac{\partial \varepsilon_{\text{eff}}}{\partial u_{yy}} = -\frac{\left(\varepsilon_{\text{eff}} - 1\right)^2}{2} \frac{1 \mp 1.297 \, pM \cos 2\phi_{K_{\text{eff}}} + 0.916 \, p^4 M^2}{pM} \qquad , \tag{7a}$$

For small $p$, the high order term $0.916 \, p^4 M^2$ is negligible, and Eq.(7a) goes back to Eq. (5a). But for large values of the filling ratio $p$, this correction term is useful for obtaining a precise value of the electrostrictive tensor. Similarly, for the hexagonal lattice, we have

$$\frac{\partial \varepsilon_{\text{eff}}}{\partial u_{xx}}, \frac{\partial \varepsilon_{\text{eff}}}{\partial u_{yy}} = -\frac{\left(\varepsilon_{\text{eff}} - 1\right)^2}{2} \frac{1 \mp 0.5 \, pM \cos 2\phi_{K_{\text{eff}}} + 0.375 \, p^6 M^2}{pM} . \tag{7b}$$

In these expressions, $p = \pi r_{\text{c}}^2 / a^2$ for square and $2\pi r_{\text{c}}^2 / \left(\sqrt{3} a^2\right)$ for hexagonal lattices, $r_{\text{c}}$ and $a$ denote the radius of discrete cylinders and lattice constant, and $\phi_{K_{\text{eff}}}$ indicates the direction of the effective wave vector, defined by macroscopic fields $\tan \phi_{K_{\text{eff}}} = \text{Re}\left[-E_x H_z^*\right] / \text{Re}\left[E_y H_z^*\right]$ in the effective media. The numerical coefficients such as 1.297, -0.596, 0.916 for the square lattice in the equations are the results of the summations of infinite series (which have analytic expressions) with 3 significant digits. The same is true for the corresponding values in the hexagonal lattice. All the details of the analytical derivation are given in the Appendix. It is important to note that the electrostrictive terms depend on the microscopic details of the underlying structure, even in the $\omega \to 0$ and $k \to 0$ and $p \to 0$ limit. In a real lattice as showed in Fig.2, the strain can change the shape of unit cell, and hence the local field, leading to a change of $\varepsilon_{\text{eff}}$. We see that $\partial \varepsilon_{\text{eff}} / \partial u_{\text{ik}}$ derived from a specific lattice contains the corresponding symmetry information of specific lattice fields. And in principle, different lattices have different lattice fields, which leads to the different values of $\partial \varepsilon_{\text{eff}} / \partial u_{\text{ik}}$, so square and hexagonal lattices have different force densities even though they can share the same values of $\varepsilon_{\text{eff}}$.

### D. Calculation of electrostrictive terms using numerical finite difference and comparison with analytic results

We checked the above analytically-derived formulae (Eqs.(5-7)) against numerical brute-force full-wave calculations. If we have a numerical algorithm that can calculate the photonic band structure, we can determine the effective permittivity by calculating the slope of the photonic band dispersion in the $\omega \to 0$ and $k \to 0$ limit. We can repeat the calculation after deforming the unit cell and the electrostrictive coefficients can be calculated using finite differences. We choose to use a commercial package COMSOL [41] to compute the photonic band structure. The lowest eigen-frequency $f_0$ of a specific unit cell with periodic boundary condition can be determined numerically when the wave vector $\vec{K}_{\text{eff}}$ is specified. In the long-wavelength limit, the effective permittivity can then be calculated with the relationship $K_{\text{eff}} = \sqrt{\varepsilon_{\text{eff}}} \left( 2\pi f_0 \right) / c$ (note that $\mu_r = 1$). We then consider an infinitesimally strained unit cell with the cylinders inside the unit remaining unchanged as shown in Fig. 2 ($\Delta a \ll a$). The magnitude of the strain tensor $u_{\text{ik}}$ can be determined according to the standard definition. For the square lattice, $u_{\text{xx}} = 2\Delta a / a$ as shown in Fig. 2(a), and the shearing strain $u_{\text{xy}} = \Delta a / a$ as shown in Fig. 2(b). For the hexagonal lattice, the stretching and shearing are given by $u_{\text{xx}} = 2\Delta a / a$ and $u_{\text{xy}} = \Delta a / \left( \sqrt{3}a \right)$ as shown respectively in Fig. 2(c) and Fig. 2(d). We repeat the numerical procedures to obtain the permittivity $\tilde{\varepsilon}_{\text{eff}}$ for the strained lattice, and the electrostrictive term can be found using finite-difference $\partial \varepsilon_{\text{eff}} / \partial u_{\text{ik}} = \left( \tilde{\varepsilon}_{\text{eff}} - \varepsilon_{\text{eff}} \right) / u_{\text{ik}}$. We compare the numerically calculated and the analytical results for the case $\varepsilon_c = 8$ in Fig. 3, and we found that agreements are excellent. In Figs. 3(a),(b),(e),(f), the radii of cylinders are $r_c = 0.3a$ in square and $r_c = 0.279a$ in hexagonal lattices respectively. These cases correspond to same filling ratio of $p_{\text{small}} = 0.2828$ and they give the same effective permittivity $\varepsilon_{\text{eff}} = 1.57$. At this filling ratio, the small filling ratio results (Eq. 5 and 6) marked by the red lines and the high filling ratio results (Eq. 7) marked by the blue lines are essentially the same and they both agree with the numerical results. Fig. 3(c),(g),(d),(h) shows the results for a large filling ratio of $p_{\text{large}} = 0.6362$ ($\varepsilon_{\text{eff}} = 3.08$) with the radius of cylinders being $r_{\text{c}} = 0.45a$ in the square and $r_{\text{c}} = 0.425a$ in the hexagonal lattice. At this high filling ratio, the corrections for high order terms in Eq.(7) are needed to obtain good agreement with the numerical results. These comparisons give us confidence that the analytically derived results are correct.

### III. COMPUTATION OF THE LIGHT-INDUCED FORCE DENSITY: LATTICE SYSTEM VS. EFFECTIVE-MEDIUM SYSTEM

In principle, we can obtain the force density distribution of the metamaterial by calculating the optical force acting on each individual dielectric cylinder inside the cylinder array. We need to compute the electromagnetic fields at each point of the whole system. To compute the optical force acting on the cylinder at the $i^{\text{th}}$ row and $j^{\text{th}}$ column in the square lattice, we draw a closed boundary that encircles that cylinder, as indicated in the inset of Fig. 1(a) where black dashed lines are shown surrounding one of the cylinder. The integral of stress tensor along the boundary as marked by dashed lines (Maxwell and Helmholtz stress tensor are the same as the boundary lines are cutting through vacuum) gives the optical force on the discrete cylinder on the site $(i, j)$.

$$\vec{f}_{\text{L}}(i, j) = \oint_{\text{C}} \vec{\vec{T}}_{\text{Maxwell}}^{\text{Lattice fields}} \cdot d\vec{l} \ , \tag{8}$$

where the sub-index "L" indicates the force is evaluated with fields of the microscopic lattice. Such lattice field computations give by definition the correct optical force acting on each cylinder in the lattice. It requires significant effort as we need to solve the multiple scattering problem of nearly 2000 cylinders in order to obtain $\vec{f}_{\text{L}}(i, j)$ through Eq. (8). On the contrary, we can easily obtain the field in effective model as we just need to solve the scattering problem for one single object with an effective permittivity. After deriving the electrostrictive terms, we can use the fields in the effective medium to compute the force density using the Helmholtz tensors on the same site $(i, j)$ to see if it agrees with $\vec{f}_{\text{L}}(i, j)$ calculated using the microscopic model. For comparison purposes, we also compute the force using the Maxwell tensor.

$$\vec{f}_{\text{M}}(i, j) = \oint_{\text{C}} \vec{\vec{T}}_{\text{Maxwell}}^{\text{Effective fields}} \cdot d\vec{l} \qquad \vec{f}_{\text{H}}(i, j) = \oint_{\text{C}} \vec{\vec{T}}_{\text{Helmholtz}}^{\text{Effective fields}} \cdot d\vec{l} \tag{9}$$

We note that the entire object is composed of 1976 discrete cylinders in Fig. 1(a), the high density of cylinders means that the force acting on individual cylinders can be used to define the force density which can be compared with the effective medium results. And then we compare the profiles of $\vec{f}_M(i,j)$ and $\vec{f}_H(i,j)$ with $\vec{f}_L(i,j)$, to identify whether $\vec{f}_M(i,j$ or $\vec{f}_H(i,j)$ gives results closer to the correct result $\vec{f}_L(i,j)$. The external light is a plane wave incident from the left side. The results of $E_z$ polarization are showed in Fig.4. The left, middle, and right columns are force density results for $\vec{f}_L(i,j)$, $\vec{f}_M(i,j)$ and $\vec{f}_H(i,j)$ respectively, and the upper row is for x components of the three kinds of force density, the lower row is for y components. For the $E_z$ polarization, only $E_z$, $H_x$, $H_y$ components are nontrivial. According to the electrostrictive terms showed in Eq.(4), the Maxwell and Helmholtz tensors in Eq.(2) and Eq.(3) are coincidently the same for this polarization. So, the Maxwell and Helmholtz tensor give the same results using effective fields and both agree with the correct force density profile $\vec{f}_L(i,j)$. On the contrary, the Maxwell tensor and Helmholtz tensor do not share the same expression for the $H_z$ polarization. The lattice fields and effective fields are not the same in this case. The force densities computed using these two tensors are different, as shown in Fig.5 and the results indicate that Maxwell tensor is obviously wrong, while the Helmholtz tensor gives force density profile that is very similar to the correct one.

## IV. THE SYMMETRY DEPENDENCE OF THE FORCE DENSITY

The results in Fig. 5 shows that Helmholtz tensor gives results that are obviously superior to the Maxwell tensor, but we still need to quantitatively investigate how accurate the Helmholtz tensor can be. We switch to a slab geometry, which is easier for quantitative comparisons. We first consider a smaller filling ratio of $p_{small} = 0.2828$ ( $\varepsilon_{eff} = 1.57$ ) for the square lattice. We arrange 50 layers of $\varepsilon_c = 8$ cylinders in the x direction, and the number of cylinders in each layer is infinite along the y direction. Such configuration can be regarded as a slab according to effective medium theories. We consider a $H_z$ polarized plane wave with an incident angle of 30 degrees, as shown schematically in the inset of Fig. 6(a). The x-component of calculated $\vec{f}_L(i,j)$, $\vec{f}_M(i,j)$ and $\vec{f}_H(i,j)$ are shown respectively by black open squares, red circles and blue triangles in the left columns of Fig.6. Figure 6(a) displays the data on a larger scale so as to show the jump of the force density on the boundary of the media, while Fig.6(b) shows the same data of Fig.6(a) on a smaller scale in order to highlight the differences (if any) of these three kinds of force density. In accordance with the results for circular domains (Fig. 5), only Helmholtz tensor gives the correct force density profile. We repeat the same calculations for the hexagonal lattice which has same $\varepsilon_{eff}$ as the square lattice and the results for the hexagonal lattice are shown in Fig.6(c) and 6(d).

Noting that the distance between the adjacent layers in the hexagonal lattice is $\sqrt{3}a/2$, we need 58 layers in x direction in order to keep same thickness of the effective slab. We note again that the Helmholtz tensor gives the correct force densities, while the Maxwell tensor fails.

We note the total force acting on the slab is decided solely by $\varepsilon_{\text{eff}}$, and it must be exactly the same for the square or hexagonal lattices as long as they share the same $\varepsilon_{\text{eff}}$. The microscopic details are irrelevant for the total force. However, the force density inside the system is determined by the Helmholtz tensor, which requires electrostrictive terms that are explicitly symmetry dependent. For the results showed in Fig.6, if the incident power along x direction is $6.25\,\text{W/m}$, then the total forces of the two lattices are both $6.71 \times 10^{-11}\,\text{N/m}$ and they are the same up to numerical errors, but the force density are different as can be observed if we compare the results of Fig.6(b) and 6(d). We also note from Fig.6(a) and 6(c) that the difference is more noticeable on the boundaries.

We now consider a large filling ratio of $p_{\text{large}} = 0.6362$ ($\varepsilon_{\text{eff}} = 3.08$) and the results are shown in Fig. 7. With the same incident power, the total forces are $4.34 \times 10^{-10}\,\text{N/m}$ for both lattices, while the force density in the square lattice is different from that in the hexagonal lattice, and the symmetry-induced difference is relatively larger than the difference in small filling ratio case. That is because the larger $\varepsilon_{\text{eff}}$ gives the larger difference of electrostrictive terms between square and hexagonal lattices as showed in Eqs.(5) and (6). However, if we observe the results of Helmholtz tensor in Fig.7(a) and 7(b) carefully, we find there is still a small difference between the Helmholtz tensor results and the force densities calculated using the microscopic lattice system. This discrepancy highlights some intrinsic limitation of using effective medium to describe optically force density which will be discussed in the next section.

## V. DISCUSSION

After showing numerical results and the comparison with analytic results, some discussions are now in order. First, we find that the failure of Maxwell tensor in describing the body forces in the $H_z$ polarization can be attributed to the depolarized fields. In the $E_z$ polarization, neither the external E field nor H field induces a depolarization field in the 2D dielectric cylinder array. So, the fields of the microscopic system and the effective fields in the homogenized medium are essentially the same along the integral path (boundary of unit cell), and the force density calculated in the microscopic lattice and the effective medium using Maxwell tensor are hence the same. For the $H_z$ polarization, the incident E field induces depolarization field in xy plane in the lattice system, and the resulting lattice fields have complex patterns due to the polarization charges on the surface of the cylinders. On the other hand, in the effective-medium system, the effective fields can be regarded as the averaged values of the lattice fields. Then according to Eq.(2)(3), we see that the force acting on each discrete cylinder is $\vec{f} \sim \left\langle E_{\text{lattice}}^2 \right\rangle$ ( here$<\dots>$ means taking a spatial average over the boundary of the unit cell), while force acting on the same region in effective model is $\vec{f} \sim \left\langle E_{\text{lattice}} \right\rangle^2$. As $E_{\text{lattice}}$ is a rapidly changing spatial function, $\left\langle E_{\text{lattice}}^2 \right\rangle \neq \left\langle E_{\text{lattice}} \right\rangle^2$, Maxwell tensor cannot be used to calculate force density. We note in particular that $E_{\text{lattice}}$ depends on the symmetry of the lattice. Once we adopt an effective medium approach, we are taking the average of the fields and the symmetry property of the fields are lost in the spatially averaged field $\left\langle E_{\text{lattice}} \right\rangle$. In order to calculate the force density based on the effective medium field, additional electrostrictive terms $\partial \varepsilon_{\text{r}} / \partial u_{\text{ik}}$ are needed to compensate for the missing information.

The importance of the electrostrictive term can be seen from another angle. If we take out the electrostrictive terms $\partial \varepsilon_{\text{r}} / \partial u_{\text{ik}}$ from the Helmholtz tensor, the remaining terms of the Helmholtz tensor will then be [7, 20]:

$$T_{\text{ik}} = \frac{1}{2} \text{Re} \left[ \varepsilon_0 \varepsilon_{\text{r}} E_{\text{i}} E_{\text{k}}^* - \frac{1}{2} \varepsilon_0 \varepsilon_{\text{r}} \delta_{\text{ik}} E^2 + \mu_0 H_{\text{i}} H_{\text{k}}^* - \frac{1}{2} \mu_0 H^2 \delta_{\text{ik}} \right] \qquad . \qquad (10)$$

Here we have taken $\mu_{\text{r}} = 1$. In isotropic dielectric media and in the absence of extraneous charge or current, it can be shown that (see Appendix B or [7])

$$\oint_S \vec{\vec{T}} \cdot \mathrm{d}\vec{s} = -\int_V \frac{1}{2} \varepsilon_0 E^2 \nabla \varepsilon_r \, \mathrm{d}\tau \qquad . \qquad (11)$$

We note that Eq. (10) is the Minkowski stress tensor for isotropic non-magnetic materials. In our dielectric system which has isotropic $\varepsilon_{\text{eff}}$, the Minkowski and Abraham tensors are the same if we are considering time-averaged forces. Eq.(11) indicates that the force density in the bulk derived from the Minkowski/Abraham tensor is zero unless the integral region $V$ is inhomogeneous. Therefore, the force density inside a homogenous object (where $\nabla \varepsilon_r = 0$) calculated by Helmholtz tensor is solely determined by the product of electrostrictive term and the square of effective E field, and the difference of the internal force density between different lattices comes from this term. We also note many authors recently favored the use of the Einstein-Laub tensor [24,26]. In the dielectric systems we are concerned with, the Helmholtz tensor is equal to the Minkowski/Abraham tensor plus additional electrostriction terms while the Einstein-Laub tensor (in the electrostatic limit) would be the Minkowski/Abraham tensor plus additional terms proportional to $\nabla(\vec{P} \cdot \vec{E})$. The Einstein-Laub tensor will not give the same result as the Helmholtz tensor. The results of Liberal *et al.* [24] show clearly that the force density calculated using Einstein-Laub tensor is completely characterized by the effective electric and magnetic susceptibilities. As such, it does not carry any information related to electrostriction or magnetostriction and the force densities calculated using the Einstein-Laub tensor, as determined by effective susceptibilities and macroscopic fields, will be insensitive to the symmetry of the underlying lattice. There is evidence that the recently proposed modified Einstein-Laub tensor [28] gives accurate results for the total force/torque for an object immersed in liquid but it still does not give electrostriction related information which originates from multiple scattering effects.

The results in Fig. 7(a) and 7(b) show that the results calculated using the Helmholtz tensor do not match exactly those of the microscopic lattice, even though the results are obviously much better than those calculated using the Maxwell tensor without electrostriction. We will see that the small discrepancy stems from the explicit dependence of the electrostrictive terms on the direction of $\vec{K}_{\text{eff}}$. If there is only one specific $\vec{K}_{\text{eff}}$ as we assumed in the derivation of electrostrictive terms in Appendix A, the explicit value of such electrostrictive term can be rigorously defined. However, if the field inside a metamaterial is the result of the interference of multiple wave vectors due to the reflection at the boundaries, the value of $\phi_{K_{\text{eff}}}$ cannot be defined unambiguously. For instance, let us consider the slab configuration as shown in Fig.6 and Fig.7. When light is incident from the left boundary at a particular incident angle, there will be a transmitted wave $\vec{K}_1 = K_x \hat{x} + K_y \hat{y}$ going from left to right inside the slab, and there is a reflected wave $\vec{K}_2 = -K_x \hat{x} + K_y \hat{y}$ going from right to left. We note again the electrostrictive terms (Eq. 5 to 7) depends explicitly on $\phi_{K_{\text{eff}}}$, which can be meaningfully defined only if there is a one-to-one correspondence between the field and one $\vec{K}_{\text{eff}}$. If the field at one point inside metamaterial comes from the interference of multiple k-vectors, the one-to-one correspondence no longer exists and the best we can do is to replace the $\vec{K}_{\text{eff}}$ by the local Poynting vector. We will show in Appendix C that it is a good approximation if the magnitude of reflected field is much smaller than the magnitude of transmitted field. If the total fields inside the metamaterial are regarded as the transmitted wave slightly perturbed by reflected waves, it can be shown that the small reflected wave just rotates the direction of the transmitted wave (Appendix C). In that case, the direction of $\vec{K}_{\text{eff}}$ can still be defined by the equation $\tan\phi_{K_{\text{eff}}} = \text{Re}\left[-E_x H_z^*\right]\big/\text{Re}\left[E_y H_z^*\right]$ inside the effective media. For example, for a small effective permittivity $\varepsilon_{\text{eff}} = 1.57$, the reflected field is indeed much smaller than the transmitted field as the impedance difference between air and the media is not large, and the results of Fig.6 show that Helmholtz tensor can give force densities that are almost the same as the results calculated using the microscopic model. This shows that such perturbation approach is good enough in this situation. In addition to the slab geometry, the results shown in Fig.4 for a circular boundary also show that the approach of using the local Poynting vector can give accurate force densities compared with those calculated using the microscopic model. However, if the impedance mismatch at boundaries is not small, as in the case of $\varepsilon_{\text{eff}} = 3.08$, the error of the perturbation calculation begins to emerge. We can see from Fig.7(b) that the Helmholtz tensor with the electrostrictive terms calculated using the local Poynting vector do not give exactly the same results as those of the microscopic structure. This is because the reflected field is not small enough compared with the transmitted field.

## VI. CONCLUSION

We considered prototypical 2D systems composing of arrays of deep sub-wavelength dielectric cylinders with lattice constants that are very small compared with the wavelength so that standard effective medium theory should provide an excellent description of the optical properties. We computed the optical force density inside this composite system and compared the corresponding results with those obtained using Helmholtz and Maxwell stress tensors within the framework of an effective medium approach in the sense that the fields are obtained using an effective $\varepsilon_{\text{eff}}$ rather than the microscopic lattice system. Our results showed that we should use the Helmholtz stress tensor to calculate optical force density inside a metamaterial. The Helmholtz stress tensor carries electrostrictive terms $\partial\varepsilon_r/\partial u_{ik}$ (and magnetostrictive terms if $\mu_r \neq 1$), which are not given in standard effective theories. Using multiple scattering theory, we succeeded in deriving analytical expressions for $\partial\varepsilon_r/\partial u_{ik}$ for both square and hexagonal lattices, and we find that $\partial\varepsilon_r/\partial u_{ik}$ depends not only explicitly on the symmetry of underlying lattice, but it also depends on the direction of a local effective wave vector. These analytic results actually have interesting implications. In the spirit of using stress tensors to compute forces within the framework of an effective medium approach, we frequently assume that if the effective medium description of the fields is highly precise, then every physical quantity can be obtained accurately. Our results show that it is not the case for optically induced body forces. Even if the fields in an effective medium description are highly precise, we still need to know the symmetry of the underlying lattice before we can apply the stress tensor, as the expressions of electrostrictive corrections depend explicitly on symmetry. Worse still, we need to map the local field to one specific wave vector, which is difficult when impedance mismatch at boundaries is big as the reflections at the boundaries would introduce multiple wave vectors at any point inside the photonic crystal or metamaterial. However, numerical calculations show that the Helmholtz stress tensor can indeed give rather satisfactory force density profiles as long as the impedance mismatch at the boundaries is not big so that a local effective wave vector can be defined. We also note that our results can be extended straightforwardly to systems requiring magnetostrictive terms $\partial \mu_r / \partial u_{ik}$.

## ACKNOWLEDGEMENTS

This work is supported by Hong Kong Research Grant Council through AoE grant AOE/P-02/12 and also by National Natural Science Foundation China (Nos. 11474057, 11174055), and HK RGC ECS 209913.

## APPENDIX A: CALCULATION OF ELECTROSTRICTIVE COEFFICIENTS

### Basic notation in multiple scattering theory (MST).

We will use the standard multiple scattering theory that has been developed for two dimensional infinite periodic systems comprising a lattice of cylinders. Details of this approach can be found in the literature. We will adopt the formalism and the notations of Ref. [39].

We start from the scattering of a single cylinder identified by subscript P positioned at $\vec{r}_P$. The cylinder is infinite along z direction. For the $E_z$ polarization, the incident and scattered field distribution can be written as,

$$E_{z}(\vec{r}_P) = E_{z,\text{inc}}(\vec{r}_P) + E_{z,\text{sca}}(\vec{r}_P) = \sum_m A_m^P J_m(k_0 r_P) e^{im\phi_P} + \sum_m B_m^P H_m^{(1)}(k_0 r_P) e^{im\phi_P} \quad (m \in \mathbb{Z}). \quad (A1)$$

Here, $J_m$ and $H_m^{(1)}$ are Bessel and first kind Hankel functions, $k_0$ stands for wave vector in vacuum, and $\vec{r}_P = (r_P, \phi_P)$ is the position vector in cylindrical coordinates of the center of the cylinder P. The incident and scattered field coefficients are $A_m^P$ and $B_m^P$. Taking into account of the electromagnetic boundary condition on the surface of cylinder, we have:

$$B_m^P = F_m A_m, \quad (A2)$$

where the Mie coefficients ($F_m$) for $E_z$ polarization are [39, 40]

$$F_{\mathrm{m}} = -\frac{J_{\mathrm{m}}(\sqrt{\varepsilon_c\mu_c}\,k_0 r_c)J_{\mathrm{m}}{}'(k_0 r_c) - \sqrt{\varepsilon_c/\mu_c}\,J_{\mathrm{m}}{}'(\sqrt{\varepsilon_c\mu_c}\,k_0 r_c)J_{\mathrm{m}}(k_0 r_c)}{J_{\mathrm{m}}(\sqrt{\varepsilon_c\mu_c}\,k_0 r_c)H_{\mathrm{m}}^{(1)}{}'(k_0 r_c) - \sqrt{\varepsilon_c/\mu_c}\,J_{\mathrm{m}}{}'(\sqrt{\varepsilon_c\mu_c}\,k_0 r_c)H_{\mathrm{m}}^{(1)}(k_0 r_c)} \quad , \tag{A3}$$

And for $H_z$ polarization, the Mie coefficients become

$$F_{\mathrm{m}} = -\frac{J_{\mathrm{m}}(\sqrt{\varepsilon_c\mu_c}\,k_0 r_c)J_{\mathrm{m}}{}'(k_0 r_c) - \sqrt{\mu_c/\varepsilon_c}\,J_{\mathrm{m}}{}'(\sqrt{\varepsilon_c\mu_c}\,k_0 r_c)J_{\mathrm{m}}(k_0 r_c)}{J_{\mathrm{m}}(\sqrt{\varepsilon_c\mu_c}\,k_0 r_c)H_{\mathrm{m}}^{(1)}{}'(k_0 r_c) - \sqrt{\mu_c/\varepsilon_c}\,J_{\mathrm{m}}{}'(\sqrt{\varepsilon_c\mu_c}\,k_0 r_c)H_{\mathrm{m}}^{(1)}(k_0 r_c)} \quad . \tag{A4}$$

Now, for a 2D periodic system, in addition to the external field, the incident field acting on cylinder P should include the field scattered by other cylinders at $\vec{r}_Q$, $\vec{r}_Q \neq \vec{r}_P$, $\vec{r}_Q = (r_Q, \phi_Q)$:

$$E_{\mathrm{z,inc}}(\vec{r}_P) = \sum_{\mathrm{m}} A_{\mathrm{m}}^P J_{\mathrm{m}}(k_0 r_P)e^{\mathrm{i}\mathrm{m}\phi_P} + \sum_{Q\neq P}\sum_{\mathrm{m'}} B_{\mathrm{m'}}^Q H_{\mathrm{m'}}^{(1)}(k_0 r_Q)e^{\mathrm{i}\mathrm{m'}\phi_Q} \qquad (\mathrm{m}\in\mathbb{Z}) \quad . \tag{A5}$$

Using Graf's additional theorem (see, e.g., Milton Abramowitz, Irene A. Stegun, *Handbook of mathematical functions with Formulas, Graphs, and Mathematical Tables*, 1972), we expand the Hankel function centered at cylinder Q in the basis of Bessel function centered at cylinder P

$$H_{\mathrm{m'}}^{(1)}(k_0 r_Q)e^{\mathrm{i}\mathrm{m'}\phi_Q} = \sum_{\mathrm{m'}}\left(H_{\mathrm{m'-m'}}^{(1)}(k_0 R_{PQ})e^{\mathrm{i}(\mathrm{m'-m'})\Phi_{PQ}}\right)J_{\mathrm{m'}}(k_0 r_P)e^{\mathrm{i}\mathrm{m'}\phi_P} \quad . \tag{A6}$$

Where $\vec{R}_{PQ} = (R_{PQ}, \Phi_{PQ}) = \vec{r}_P - \vec{r}_Q$ is the vector that directs from cylinder P to cylinder Q. In addition, we impose the Bloch condition

$$B_{\mathrm{m'}}^Q = B_{\mathrm{m'}}^P\, e^{\mathrm{i}\vec{K}_{\mathrm{eff}}\cdot\vec{R}_{PQ}} \quad . \tag{A7}$$

Here, $\vec{K}_{\mathrm{eff}}$ is the effective propagating wave vector, and we can obtain the effective permittivity from to this parameter in dielectric system (with permeability equal to 1):

$$\varepsilon_{\mathrm{eff}} = \left(K_{\mathrm{eff}}/k_0\right)^2 \quad . \tag{A8}$$

We substitute Eq.(A6) and Eq.(A7) into Eq.(A1), the second part on the right side is:

$$\sum_{Q\neq P}\sum_{\mathrm{m'}} B_{\mathrm{m'}}^Q H_{\mathrm{m'}}^{(1)}(k_0 r_Q)e^{\mathrm{i}\mathrm{m'}\phi_Q} = \sum_{Q\neq P}\sum_{\mathrm{m'}} B_{\mathrm{m'}}^P\, e^{\mathrm{i}\vec{K}_{\mathrm{eff}}\cdot\vec{R}_{PQ}}\sum_{\mathrm{m'}}\left(H_{\mathrm{m'-m'}}^{(1)}(k_0 R_{PQ})e^{\mathrm{i}(\mathrm{m'-m'})\Phi_{PQ}}\right)J_{\mathrm{m'}}(k_0 r_P)e^{\mathrm{i}\mathrm{m'}\phi_P} \quad .$$

If we define the lattice sum $S_{\mathrm{m'-m}}$ in real space:

$$S_{\mathrm{m'-m}} = \sum_{Q\neq P} H_{\mathrm{m'-m}}^{(1)}(k_0 R_{PQ})e^{-\mathrm{i}(\mathrm{m'-m})\Phi_{PQ}}e^{\mathrm{i}\vec{K}_{\mathrm{eff}}\cdot\vec{R}_{PQ}} \qquad , \qquad S_{\mathrm{m'-m}} = -\left(S_{\mathrm{m-m'}}\right)^* \quad . \tag{A9}$$

Eq.(A5) can be simplified as:

$$E_{\mathrm{z,inc}}(\vec{r}_P) = \sum_{\mathrm{m}} A_{\mathrm{m}}^P J_{\mathrm{m}}(k_0 r_P)e^{\mathrm{i}\mathrm{m}\phi_P} + \sum_{\mathrm{m}}\sum_{\mathrm{m'}} B_{\mathrm{m}}^P S_{\mathrm{m'-m}} J_{\mathrm{m'}}(k_0 r_P)e^{\mathrm{i}\mathrm{m'}\phi_P} \qquad . \tag{A10}$$

The lattice sum can be transformed to the summation in reciprocal space not only for easier analytical operation, but also for faster numerical convergence. All details can be found elsewhere [39,40]:

$$S_n = \frac{1}{J_{n+1}(k_0 a)} \left[ \frac{4 i^{n+1} k_0}{\Omega} \sum_{K_h} \frac{J_{n+1}(Q_h a)}{Q_h (k_0^2 - Q_h^2)} e^{in\phi_{Q_h}} - \left( H_1(k_0 a) + \frac{2i}{\pi k_0 a} \right) \delta_{n0} \right]_{(n \geq 0)}, \quad S_{-n} = -\left( S_n \right)^*. \text{(A11)}$$

Here, $r_c$ and $a$ denote the radius of cylinders and the lattice constant. $\vec{Q}_h = \vec{K}_{eff} + \vec{K}_h$, where $K_h$, $\Omega$ denotes the reciprocal vector and the area of unit cell. We find that the lattice sum is highly dependent on the geometry of structure.

Applying the electromagnetic boundary condition, we obtain the secular equation based on Eq.(A2):

$$\sum_{m'} \left( F_m S_{m'-m} - \delta_{m'm} \right) B_{m'} = -F_m A_m. \tag{A12}$$

To obtain $\vec{K}_{eff}$, we solve the secular equation :

$$\det \left| F_m S_{m-m'} - \delta_{mm'} \right| = 0. \tag{A13}$$

We consider systems in which the cylinders are made with a simple dielectric material, characterized by a $\varepsilon_c$, which is a finite constant number, and $\mu_c = 1$. We consider the long wavelength limit: $k_0 a < K_{eff} a \ll 1 < K_h a$. In that limit, $\vec{K}_{eff}$ and $\vec{K}_h$ in $S_n$ can be decoupled as:

$$S_n = \frac{4 i^{n+1}}{k_0^2 \Omega} \frac{\varepsilon_{eff}^{n/2} e^{-in\phi_{K_{eff}}}}{1 - \varepsilon_{eff}} - \frac{2^{n+3} i^{n+1} (n+1)!}{k_0^n a^{n-2} \Omega} \Gamma_n. \tag{A14}$$

where

$$\Gamma_n = \sum_{K_h \neq 0} \frac{J_{n+1}(K_h a)}{(K_h a)^3} e^{-in\phi_{K_h}} \tag{A15}$$

***Using perturbation calculation to obtain electrostrictive coefficients***

*(i) For $E_z$ polarization:*

$$F_0 = \frac{i\pi}{4} (\varepsilon_c - 1)(k_0^2 r_c^2) \quad , \qquad F_m = F_{-m} = 0 \big|_{m>0}, \text{(in the long wavelength limit)}$$

The secular equation is simply

$$F_0 S_0 - 1 = 0.$$

Obviously, the lattice sum $S_0$ is dominated by the first term in the right hand side of Eq.(A11). (in the limit $k_0 a < K_{eff} a \ll 1 < K_h a$). Take the square lattice for instance, when the deformation of unit cell is characterized by the strain tensor, only the stretching (diagonal terms) affects the value of $S_0$ by the perturbation in the value of unit cell size $\Omega = a^2 \left( 1 + u_{xx} \right)$. Consequently, it is easy to obtain the results showed in Eq. (4).

*(ii) For $H_z$ polarization:*

$$F_0 = 0 \quad , \qquad F_m = F_{-m} = \frac{\pi i}{2^{2m} m!(m-1)!} \frac{\varepsilon_c - 1}{\varepsilon_c + 1} (k_0 r_c)^{2m} \bigg|_{m>0} \quad .$$

As the reciprocal vector $\vec{K}_h$ depends on the specific geometry of lattice structure, the perturbation of parameter $\Gamma_n$ should also be taken into consideration as shown in the following.

### Stretching in the square lattice (Fig. 2(a))

The secular equation up to quadrupole precision is (the elements that are not contributing to the determinant are not showed here)

$$\begin{vmatrix} -1 & & & & F_{-3}S_{-4} & & \\ & -1 & & & & F_{-2}S_{-4} & \\ & & F_{-1}S_0-1 & & F_{-1}S_{-2} & & F_{-1}S_{-4} \\ & & & -1 & & & \\ F_1S_4 & & F_1S_2 & & F_1S_0-1 & & \\ & F_2S_4 & & & & -1 & \\ & F_3S_4 & & & & & -1 \end{vmatrix}$$

$$= \left(1 - F_2^2 S_4^2\right)\left(F_1 S_0 - iF_1\left|S_2\right| + F_1 F_3 \left|S_4\right|^2 - 1\right)\left(F_1 S_0 + iF_1\left|S_2\right| + F_1 F_3 \left|S_4\right|^2 - 1\right)$$

$$= 0.$$

Only the middle term can be zero, so that if we want to include the quadrupole contributions for higher precision, the secular equation to that order is

$$F_1 S_0 - iF_1\left|S_2\right| - F_1 F_3 \left|S_4\right|^2 - 1 = 0 \tag{A16}$$

Therefore, we have to figure out the perturbation terms in $S_2$ and $S_4$

Define $(h_i, h_j) \in \mathbb{Z}$ , $\vec{K}_h = h_i \frac{2\pi}{a}\hat{x} + h_j \frac{2\pi}{a}(1-u_{xx})\hat{y}$ , $\Omega = a^2\left(1 + u_{xx}\right)$ , $R_K = \sqrt{h_i^2 + h_j^2}$ ,

$\cos\psi = h_i / R_K$ , $M = \frac{\varepsilon_c - 1}{\varepsilon_c + 1}$

Firstly,

$$\Gamma_2 = \sum_{K_h \neq 0} \frac{J_3(K_h a)}{\left(K_h a\right)^3} e^{-2i\phi_{K_h}} \quad ,$$

where

$$e^{-2i\phi_{K_h}} = \cos 2\phi_{K_h} - i\sin 2\phi_{K_h}$$

$$= 2\cos^2\phi_{K_h} - 1 - 2i\cos\phi_{K_h}\sin\phi_{K_h}$$

$$= 2\frac{h_i^2(1-u_{xx})^2}{h_i^2(1-u_{xx})^2 + h_j^2} - 1 - 2i\frac{h_i(1-u_{xx})h_j}{h_i^2(1-u_{xx})^2 + h_j^2}$$

$$= 2\frac{h_i^2(1-2u_{xx})(1+2\frac{h_i^2}{h_i^2+h_j^2}u_{xx})}{h_i^2+h_j^2} - 1 - 2i\frac{h_ih_j(1-u_{xx})(1+2\frac{h_i^2}{h_i^2+h_j^2}u_{xx})}{h_i^2+h_j^2}$$

$$= 2\cos^2\psi\left[1 - 2\sin^2\psi\, u_{xx}\right] - 1 - 2i\cos\psi\sin\psi\left[1 + (\cos^2\psi - \sin^2\psi)u_{xx}\right]$$

$$= e^{-2i\psi}\left(1 - i\sin 2\psi\, u_{xx}\right)$$

Then

$$\Gamma_2 = \sum_{R_K \neq 0} \frac{J_3\left(2\pi R_K(1-\cos^2\psi\, u_{xx})\right)}{\left(2\pi R_K(1-\cos^2\psi\, u_{xx})\right)^3} e^{-2i\psi}\left(1 - i\sin 2\psi\, u_{xx}\right)$$

$$= \frac{1}{(2\pi)^3}\sum_{R_K \neq 0} R_K^{-3} e^{-2i\psi}\left[J_3(2\pi R_K) - 2\pi R_K\cos^2\psi\, J_3{}'(2\pi R_K)u_{xx}\right](1 + 3\cos^2\psi\, u_{xx})\left(1 - i\sin 2\psi\, u_{xx}\right)$$

Considering the symmetry of square lattice, we have

$$\frac{1}{(2\pi)^3}\sum_{R_K \neq 0} R_K^{-3} J_3(2\pi R_K)\, e^{-2i\psi} = 0 .$$

Thus

$$\Gamma_2 = \xi_{2xx} u_{xx} .$$

And the parameter $\xi_{2xx}$ is converged to an exclusive number for square lattice

$$\xi_{2xx} = \sum_{R_K \neq 0} \frac{(3\cos^2\psi - i\sin 2\psi)J_3(2\pi R_K) - 2\pi R_K\cos^2\psi\, J_3{}'(2\pi R_K)}{(2\pi R_K)^3} e^{-2i\psi} = 0.02684 .$$

Second, similar procedure to $\Gamma_4$

$$\Gamma_4 = \sum_{K_h \neq 0} \frac{J_5(K_h a)}{(K_h a)^3} e^{-4i\phi_{K_h}} ,$$

where

$$e^{-4i\phi_{K_h}} = \left(e^{-2i\phi_{K_h}}\right)^2 = e^{-4i\psi}\left(1 - 2i\sin 2\psi\, u_{xx}\right) ,$$

Then

$$\Gamma_4 = \sum_{R_K \neq 0} \frac{J_5\left(2\pi R_K\left(1 - \cos^2\psi u_{xx}\right)\right)}{\left(2\pi R_K\left(1 - \cos^2\psi u_{xx}\right)\right)^3} e^{-4i\psi}\left(1 - 2i\sin 2\psi u_{xx}\right)$$

$$= \frac{1}{\left(2\pi\right)^3} \sum_{R_K \neq 0} R_K^{-3} e^{-4i\psi}\left[J_5\left(2\pi R_K\right) - 2\pi R_K \cos^2\psi J_5{}'\left(2\pi R_K\right)u_{xx}\right]\left(1 + 3\cos^2\psi u_{xx}\right)\left(1 - 2i\sin 2\psi u_{xx}\right)$$

$$= \sum_{R_K \neq 0} \frac{J_5\left(2\pi R_K\right)}{\left(2\pi\right)^3 R_K^3} e^{-4i\psi} + \xi_{4xx} u_{xx}.$$

This time, the first term in $\Gamma_4$ is not zero as showed in $\Gamma_2$

$$\sum_{R_K \neq 0} \frac{J_5\left(2\pi R_K\right)}{\left(2\pi\right)^3 R_K^3} e^{-4i\psi} = 0.006269.$$

And the coefficient of the second term in $\Gamma_4$ is:

$$\xi_{4xx} = \sum_{R_K \neq 0} \frac{\left(3\cos^2\psi - 2i\sin 2\psi\right)J_5(2\pi R_K) - 2\pi R_K \cos^2\psi J_5{}'(2\pi R_K)}{\left(2\pi R_k\right)^3} e^{-4i\psi} = -0.006066.$$

So

$$\Gamma_4 = 0.006269 + \xi_{4xx} u_{xx}.$$

Substitute $\Gamma_2$ into $S_2$, and $\Gamma_4$ into $S_4$

$$S_2 = \left(1 - u_{xx}\right)\left[\frac{-4i}{k_0^2 a^2}\frac{\varepsilon_{\text{eff}}}{1 - \varepsilon_{\text{eff}}}e^{-2i\phi_{K_{\text{eff}}}} + \frac{2^5 i3!}{k_0^2 a^2}\xi_{2xx} u_{xx}\right],$$

$$S_4 = -(1 - u_{xx})\frac{2^7 i!5!}{k_0^4 a^4}\left(0.006269 + \xi_{4xx} u_{xx}\right).$$

Then

$$\left|S_2\right| = \left(1 - u_{xx}\right)\left[\frac{4}{k_0^2 a^2}\frac{\varepsilon_{\text{eff}}}{\varepsilon_{\text{eff}} - 1} + \frac{192}{k_0^2 a^2}\xi_{2xx}\cos 2\phi_{K_{\text{eff}}} u_{xx}\right].$$

And the components in Eq.(A16) are

$$F_1 S_0 = \frac{i\pi}{4}\frac{\varepsilon_c - 1}{\varepsilon_c + 1}k_0^2 r_c^2\frac{4i}{k_0^2 a^2}\frac{1}{1 - \varepsilon_{\text{eff}}}\left(1 - u_{xx}\right) = pM\frac{1}{\varepsilon_{\text{eff}} - 1}\left(1 - u_{xx}\right) \quad ,$$

$$iF_1\left|S_2\right| = \frac{i\pi}{4}\frac{\varepsilon_c - 1}{\varepsilon_c + 1}k_0^2 r_c^2 i\left(1 - u_{xx}\right)\left[\frac{4}{k_0^2 a^2}\frac{\varepsilon_{\text{eff}}}{\varepsilon_{\text{eff}} - 1} - \frac{192}{k_0^2 a^2}\xi_{2xx}\cos 2\phi_{K_{\text{eff}}} u_{xx}\right]$$

$$= -pM\left[\frac{\varepsilon_{\text{eff}}}{\varepsilon_{\text{eff}} - 1} + 1.297\cos 2\phi_{K_{\text{eff}}} u_{xx}\right]\left(1 - u_{xx}\right) \quad ,$$

$$= -pM\left[\frac{\varepsilon_{\text{eff}}}{\varepsilon_{\text{eff}} - 1} - \left(\frac{\varepsilon_{\text{eff}}}{\varepsilon_{\text{eff}} - 1} - 1.297\cos 2\phi_{K_{\text{eff}}}\right)u_{xx}\right]$$

$$F_1 F_3 \left| S_4 \right|^2 = \frac{\pi i}{4} (k_0 r_c)^2 \frac{\pi i}{768} (k_0 r_c)^6 M^2 \left| -(1-u_{xx}) \frac{2^7 i^1 5!}{k_0^4 a^4} \left( 0.006269 + \xi_{4xx} u_{xx} \right) \right|^2$$

$$= -\frac{(128 \times 120)^2}{4 \times 768 \pi^2} p^4 M^2 \left[ \left( 0.006269 + \xi_{4xx} u_{xx} \right) (1 - u_{xx}) \right]^2$$

$$= -p^4 M^2 \left[ 0.3058 + \frac{(128 \times 120)^2}{4 \times 768 \pi^2} 2 \times 0.006269 \times \left( \xi_{4xx} - 0.006269 \right) u_{xx} \right]$$

$$= -p^4 M^2 \left( 0.3058 - 1.222 u_{xx} \right).$$

Thus, the relation between $\varepsilon_{\text{eff}}$ and $u_{xx}$ can be expressed as

$$pM \frac{\varepsilon_{\text{eff}} + 1}{\varepsilon_{\text{eff}} - 1} + 0.3058 p^4 M^2 - \left[ pM \left( \frac{\varepsilon_{\text{eff}} + 1}{\varepsilon_{\text{eff}} - 1} - 1.297 \cos 2\phi_{K_{\text{eff}}} \right) + 1.222 p^4 M^2 \right] u_{xx} = 1 \qquad (A17)$$

We note that without the perturbation term due to deformation, the equation becomes

$$pM \frac{\varepsilon_{\text{eff}} + 1}{\varepsilon_{\text{eff}} - 1} + 0.3058 p^4 M^2 = 1,$$

which is exactly the famous Rayleigh mixing formula to determine effective permittivity that takes into consideration of higher filling ratio contributions.

After some algebra, we have Eq.(7a):

$$\frac{\partial \varepsilon_{\text{eff}}}{\partial u_{xx}} = -\frac{(\varepsilon_{\text{eff}} - 1)^2}{2} \frac{1 - 1.297 \, pM \cos 2\phi_{K_{\text{eff}}} + 0.916 p^4 M^2}{pM}.$$

While in the limit of small filling ratio, the high orders $p^4 M^2$ can be neglected, the electrostrictive term is simplified as a function of effective wave vector angle $\phi_{K_{\text{eff}}}$, and also depends on the macroscopic permittivity $\varepsilon_{\text{eff}}$ and symmetry of square lattice (as characterized by the number 1.297):

$$\frac{\partial \varepsilon_{\text{eff}}}{\partial u_{xx}} = -\frac{\varepsilon_{\text{eff}}^2 - 1}{2} + 1.297 \frac{(\varepsilon_{\text{eff}} - 1)^2}{2} \cos 2\phi_{K_{\text{eff}}}.$$

And then replacing $\phi_{K_{\text{eff}}}$ by $\phi_{K_{\text{eff}}} + \pi/2$, we can obtain the term $\frac{\partial \varepsilon_{\text{eff}}}{\partial u_{yy}}$. Similar calculations can be performed for other three cases as displayed in Fig. 2(b-d).

***Shearing in the square lattice (Fig. 2(b))***

We introduce a shear $u_{xy}$, then

$$\vec{K}_h = \frac{2\pi}{a} h_i \hat{i} + \frac{2\pi}{a}(h_j - 2u_{xy}h_i)\hat{j}, \Omega = a^2,$$

and

$$\Gamma_2 = \xi_{2xy} u_{xy}$$

$$\xi_{2xy} = \sum_{R_K \neq 0} \frac{J_3(2\pi R_K)(3\sin 2\psi + 4i\cos^2\psi) - 2\pi R_K \sin 2\psi J_3{}'(2\pi R_K)}{(2\pi R_K)^3} e^{-2i\psi} = 0.001243i.$$

The relation between $\varepsilon_{\text{eff}}$ and $u_{xy}$ (no high order contribution in the coefficient of $u_{xy}$) is

$$pM \frac{\varepsilon_{\text{eff}}+1}{\varepsilon_{\text{eff}}-1} + 0.3058 p^4 M^2 - pM\left(0.596\sin 2\phi_{K_{\text{eff}}} u_{xy}\right) = 1.$$

### *Stretching in the hexagonal lattice (Fig. 2(c))*

The secular function with up to 6$^{\text{th}}$ order truncation in $m$ can be written as:

$$F_1 S_0 - iF_1 |S_2| - F_1 F_5 |S_6|^2 - 1 = 0.$$

And

$$\vec{K}_h = \frac{4\pi}{\sqrt{3}a}\left[h_i\left(\frac{\sqrt{3}}{2}(1-u_{xx})\hat{i} - \frac{1}{2}\hat{j}\right) + h_j\hat{j}\right], \Omega = \frac{\sqrt{3}}{2}a^2(1+u_{xx}), R_K = \sqrt{h_i^2 - h_i h_j + h_j^2}, \cos\psi = \frac{\sqrt{3}h_i}{2R_K}.$$

Then

$$\Gamma_2 = \xi_{2xx} u_{xx}$$

$$\xi_{2xx} = \left(\frac{\sqrt{3}}{4\pi}\right)^3 \sum_{R_K \neq 0} R_K^{-3} e^{-2i\psi}\left[\left(3\cos^2\psi - i\sin 2\psi\right)J_3\left(\frac{4\pi R_K}{\sqrt{3}}\right) - \frac{4\pi R_K}{\sqrt{3}}\cos^2\psi J_3{}'\left(\frac{4\pi R_K}{\sqrt{3}}\right)\right]$$
$$= 0.0104057.$$

$$\Gamma_6 = \left(\frac{\sqrt{3}}{4\pi}\right)^3 \sum_{R_K \neq 0} J_7(2\pi R_K) R_K^{-3} e^{-6i\psi} + \xi_{6xx} u_{xx} = -0.004810 + \xi_{6xx} u_{xx}.$$

$$\xi_{6xx} = \left(\frac{\sqrt{3}}{4\pi}\right)^3 \sum_{R_K \neq 0} R_K^{-3} e^{-6i\psi}\left[\left(3\cos^2\psi - 3i\sin 2\psi\right)J_7\left(\frac{4\pi R_K}{\sqrt{3}}\right) - \frac{4\pi R_K}{\sqrt{3}}\cos^2\psi J_7{}'\left(\frac{4\pi R_K}{\sqrt{3}}\right)\right]$$
$$= 0.0095537.$$

The relation between $\varepsilon_{\text{eff}}$ and $u_{xx}$ is

$$pM \frac{\varepsilon_{\text{eff}}+1}{\varepsilon_{\text{eff}}-1} + 0.0754 p^6 M^2 - \left[pM\left(\frac{\varepsilon_{\text{eff}}+1}{\varepsilon_{\text{eff}}-1} - 0.4997\cos 2\phi_{K_{\text{eff}}}\right) + 0.4504 p^6 M^2\right]u_{xx} = 1.$$

### Shearing in the hexagonal lattice (Fig. 2(d))

We introduce a shear $u_{xy}$, then

$$\vec{K}_h = \frac{4\pi}{\sqrt{3}a}\left[ h_i\left( \frac{\sqrt{3}}{2}\hat{i} - \frac{1}{2}(1 + 2\sqrt{3}u_{xy})\hat{j}\right) + h_j\hat{j}\right], \Omega = \frac{\sqrt{3}}{2}a^2.$$

and

$$\Gamma_2 = \xi_{2xy}u_{xy}$$

$$\xi_{2xy} = \left(\frac{\sqrt{3}}{4\pi}\right)^3 \sum_{R_K \neq 0} R_K^{-3} e^{-2i\psi}\left[ J_3\left(\frac{4\pi}{\sqrt{3}}R_K\right)\left(3\sin 2\psi + 4i\cos^2\psi\right) - \frac{4\pi}{\sqrt{3}}R_K\sin 2\psi J_3{}'\left(\frac{4\pi}{\sqrt{3}}R_K\right)\right]$$
$$= -0.0208152i$$

The relation between $\varepsilon_{eff}$ and $u_{xy}$ (no high order contribution in the coefficient of $u_{xy}$) can be written as

$$pM\frac{\varepsilon_{eff} + 1}{\varepsilon_{eff} - 1} + 0.0754 p^6 M^2 - pM\left(0.999\sin 2\phi_{K_{eff}} u_{xy}\right) = 1.$$

### APPENDIX B: STRESS TENSOR WITHOUT ELECTROSTRICTIVE CORRECTIONS

The local force density acting on extraneous charge and extraneous current can be written as:

$$\vec{f}_{ex} = \rho_{ex}\vec{E} + j_{ex} \times \vec{B} = \left(\nabla \cdot \vec{D}\right)\vec{E} + \left(\nabla \times \vec{H} - \frac{\partial \vec{D}}{\partial t}\right) \times \vec{B},$$

where

$$\frac{\partial \vec{D}}{\partial t} \times \vec{B} = \frac{\partial\left(\vec{D} \times \vec{B}\right)}{\partial t} - \vec{D} \times \frac{\partial \vec{B}}{\partial t} = \frac{\partial\left(\vec{D} \times \vec{B}\right)}{\partial t} + \vec{D} \times \left(\nabla \times \vec{E}\right)$$

Then

$$f_{ex} = \left(\nabla \cdot \vec{D}\right)\vec{E} - \vec{D} \times \left(\nabla \times \vec{E}\right) + \left(\nabla \cdot \vec{B}\right)\vec{H} - \vec{B} \times \left(\nabla \times \vec{H}\right) - \frac{\partial}{\partial t}\left(\vec{D} \times \vec{B}\right)$$

We know that:

$$\nabla\left(\vec{D} \cdot \vec{E}\right) = \vec{D} \times \left(\nabla \times \vec{E}\right) + \vec{E} \times \left(\nabla \times \vec{D}\right) + \left(\vec{D} \cdot \nabla\right)\vec{E} + \left(\vec{E} \cdot \nabla\right)\vec{D}$$

Where

$$\vec{E} \times \left(\nabla \times \vec{D}\right) = \vec{E} \times \left(\nabla \times \left(\varepsilon_0\varepsilon_r\vec{E}\right)\right) = \vec{E} \times \left(\varepsilon_0\nabla\varepsilon_r \times \vec{E}\right) + \varepsilon_0\varepsilon_r\vec{E} \times \left(\nabla \times \vec{E}\right)$$

$$= \vec{D} \times \left(\nabla \times \vec{E}\right) + \varepsilon_0 E^2\nabla\varepsilon_r - \vec{E}\left(\vec{E} \cdot \varepsilon_0\nabla\varepsilon_r\right)$$

And

$$\left(\vec{E}\cdot\nabla\right)\vec{D}=\left(\vec{E}\cdot\nabla\right)\left(\varepsilon_0\varepsilon_r\vec{E}\right)=\left(\vec{E}\cdot\varepsilon_0\nabla\varepsilon_r\right)\vec{E}+\varepsilon_0\varepsilon_r\left(\vec{E}\cdot\nabla\right)\vec{E}$$

So

$$\nabla\left(\vec{D}\cdot\vec{E}\right)=2\vec{D}\times\left(\nabla\times\vec{E}\right)+\varepsilon_0 E^2\nabla\varepsilon_r+2\left(\vec{D}\cdot\nabla\right)\vec{E}$$

Then

$$\left(\nabla\cdot\vec{D}\right)\vec{E}-\vec{D}\times\left(\nabla\times\vec{E}\right)=\left(\nabla\cdot\vec{D}\right)\vec{E}+\left(\vec{D}\cdot\nabla\right)\vec{E}-\frac{1}{2}\left[\nabla\left(\vec{D}\cdot\vec{E}\right)-\varepsilon_0 E^2\nabla\varepsilon_r\right]$$

Similarly

$$\left(\nabla\cdot\vec{B}\right)\vec{H}-\vec{B}\times\left(\nabla\times\vec{H}\right)=\left(\nabla\cdot\vec{B}\right)\vec{H}+\left(\vec{B}\cdot\nabla\right)\vec{H}-\frac{1}{2}\left[\nabla\left(\vec{B}\cdot\vec{H}\right)-\mu_0 H^2\nabla\mu_r\right]$$

So in time harmonic fields, the term $\dfrac{\partial}{\partial t}\left(\vec{D}\times\vec{B}\right)$ can be eliminated after time averaging, and we can define a stress tensor:

$$T_{\mathrm{M},ik}=\frac{1}{2}\mathrm{Re}\left[\varepsilon_0\varepsilon_r E_i E_k^*-\frac{1}{2}\varepsilon_0\varepsilon_r E^2\delta_{ik}+\mu_0\mu_r H_i H_k^*-\frac{1}{2}\mu_0\mu_r H^2\delta_{ik}\right].$$

We note that the tensor is the Minkowski tensor for dielectric materials, and so we label it with "M". Then the force originating from extraneous charge and current can be written as:

$$\vec{F}_{\mathrm{ex}}=\int_V\vec{f}_{\mathrm{ex}}\mathrm{d}\tau=\oint_S\vec{T}_M\cdot\mathrm{d}\vec{s}+\int_V\frac{1}{2}\left(\varepsilon_0 E^2\nabla\varepsilon_r+\mu_0 H^2\nabla\mu_r\right)\mathrm{d}\tau\ .$$

For a dielectric system without extraneous charge or current, we have

$$\oint_S\vec{T}_M\cdot\mathrm{d}\vec{s}=-\int_V\frac{1}{2}\varepsilon_0 E^2\nabla\varepsilon_r\ \mathrm{d}\tau\ .$$

And this equation is valid for any integral path.

## APPENDIX C: $\vec{K}_{\mathrm{eff}}$ IN THE PRESENCE OF REFLECTION

Suppose that there are two specific wave vectors $\vec{K}_1=K_x\hat{x}+K_y\hat{y}$ and $\vec{K}_2=-K_x\hat{x}+K_y\hat{y}$ propagating inside a slab, where $\vec{K}_1$ is the transmitted wave through the left boundary, and $\vec{K}_2$ is the first reflected wave from right boundary. If $\varepsilon_{\mathrm{eff}}$ is small, the magnitude of reflected wave is much smaller than the magnitude of transmitted wave, and other high order reflected waves can be neglected. In that case, we have $\left|E_2\right|=\eta\left|E_1\right|$ and $\eta\ll 1$, where $E_1$ is the electric field of transmitted wave, and $E_2$ is the reflected wave, and we can write

$$E_1 = E_0 e^{iK_x x} \left( -\sin\phi_{K_1} \hat{x} + \cos\phi_{K_1} \hat{y} \right)$$

$$E_2 = \eta E_0 e^{-iK_x x} \left( \sin\phi_{K_1} \hat{x} + \cos\phi_{K_1} \hat{y} \right),$$

where $E_0$ is the magnitude of $E_1$, and the common term $e^{iK_y y}$ is not shown explicitly here.

The magnetic fields are:

$$H_1 = \sqrt{\varepsilon_0/\mu_0} E_0 e^{iK_x x} \hat{z}$$

$$H_2 = \eta \sqrt{\varepsilon_0/\mu_0} E_0 e^{-iK_x x} \hat{z}.$$

Then, the time-averaged local Poynting vector according to the local total fields is:

$$\left\langle \vec{S}_p \right\rangle = \frac{1}{2}\mathrm{Re}\left[ \vec{E} \times \vec{H}^* \right] = \frac{1}{2}\mathrm{Re}\left[ \vec{E}_1 \times \vec{H}_1^* + \vec{E}_2 \times \vec{H}_1^* + \vec{E}_1 \times \vec{H}_2^* + \vec{E}_2 \times \vec{H}_2^* \right].$$

As $\vec{E}_2 \times \vec{H}_2^*$ is second order in $\eta$, so if we only consider up to the first order in the small parameter $\eta$, we have:

$$\left\langle \vec{S}_p \right\rangle = \frac{1}{2}\mathrm{Re}\left[ \vec{E}_1 \times \vec{H}_1^* + \vec{E}_2 \times \vec{H}_1^* + \vec{E}_1 \times \vec{H}_2^* \right]$$

$$= \frac{1}{2}\sqrt{\frac{\varepsilon_0}{\mu_0}} E_0^2 \left( \cos\phi_{K_1} \hat{x} + \sin\phi_{K_1} \hat{y} \right) + \frac{1}{2}\eta\sqrt{\frac{\varepsilon_0}{\mu_0}} E_0^2 \left( e^{2iK_x x} + e^{-2iK_x x} \right)\left( \cos\phi_{K_1} \hat{x} - \sin\phi_{K_1} \hat{y} \right)$$

$$= \frac{1}{2}\sqrt{\frac{\varepsilon_0}{\mu_0}} E_0^2 \left[ \cos\phi_{K_1} \left( 1 + 2\eta\cos\left( 2K_x x \right) \right) \hat{x} + \left( \sin\phi_{K_1} \left( 1 - 2\eta\cos\left( 2K_x x \right) \right) \right) \hat{y} \right].$$

For time-averaged time harmonic fields, if there is a single propagating $\vec{K}$, the directions of $\vec{K}$ and the Poynting vector $\vec{S}_p$ are parallel. However, if there are two $\vec{K}$ propagating and interfering, $\vec{K}_{\mathrm{eff}}$ is not well defined. But we need a $\vec{K}_{\mathrm{eff}}$ to define the electrostrictive terms. For this $\eta \ll 1$ case, we still assume the local direction of $\vec{K}_{\mathrm{eff}}$ is parallel to the Poynting vector, we have

$$\tan\phi_{K_{\mathrm{eff}}} = \tan\phi_{S_p} = \frac{\mathrm{Re}\left[ -E_x H_z^* \right]}{\mathrm{Re}\left[ E_y H_z^* \right]} = \frac{\sin\phi_{K_1}\left( 1 - 2\eta\cos\left( 2K_x x \right) \right)}{\cos\phi_{K_1}\left( 1 + 2\eta\cos\left( 2K_x x \right) \right)} \approx \tan\phi_{K_1}\left( 1 - 4\eta\cos\left( 2K_x x \right) \right).$$

Therefore, at least in the situation that the magnitude of E field in $\vec{K}_2$ is much smaller that the magnitude of E field in $\vec{K}_1$, the direction of the $\vec{K}_{\mathrm{eff}}$ as we defined here can be approximately regarded as a perturbed direction according to the dominate $\vec{K}_1$ mode.

*corresponding author: phchan@ust.hk

**Figure captions**:

FIG. 1. (color online) Dielectric cylinders with relative permittivity $\varepsilon_c$ are arranged in square (panel (a)) and hexagonal lattices (panel (b)) and the array of cylinders collectively form a circular domain that is big compared with the wavelength. The system can be viewed as a finite-sized 2D photonic crystal with a circular boundary, and the corresponding unit cells are highlighted by black dashed lines in the insets which show enlarged view of the interior of the photonic crystal, exposing the details of the arrangement. The lattice constants of the photonic crystals are very small compared with wavelength so that the circular domain can be regarded as a homogenous cylinder with permittivity $\varepsilon_{\mathrm{eff}}$, and the identical square dashed line circles the same area of unit cell at the same coordinate in lattice, as well as the hexagonal lattice. In the numerical calculations, the big cylinder shown in Fig. 1(a) and Fig. 1(b) contain respectively 1976 and 1971 small cylinders. The lattice constant of the photonic crystals in Fig. 1(a) and Fig. 1(b) are adjusted so that they give the same $\varepsilon_{\mathrm{eff}}$ so that we have two systems that are the same if we use macroscopic constitutive parameters to describe the optical properties but the microscopic structures are different.

FIG. 2. (color online) This figure shows the unit cell deformations used to calculate the electrostrictive tensor components using finite difference. The upper (lower) panels are for the square (hexagonal) lattice. In (a), (b), the original square lattice unit cell is shown as semitransparent green squares. The unit cell is stretched or sheared $\Delta a$ in x direction with the deformed cell showed by blue parallelograms. Here $a$, $\phi_{K_{\mathrm{eff}}}$ and $r_{\mathrm{c}}$ denotes the side-length of cell, the direction of eigen-propagating mode $\vec{K}_{\mathrm{eff}}$ and the radius of cylinder, respectively. The black arrows show the directions to the neighbor cylinders. Panels (c) and (d) show the counterparts for the hexagonal lattice.

FIG. 3. (color online) Analytical and numerical calculated electrostrictive tensor components are compared and plotted as functions of the wave vector direction $\phi_{K_{\mathrm{eff}}}$ (see text for definition). The relative permittivity of the cylinders in the photonic crystal is $\varepsilon_{\mathrm{c}} = 8$. Results for square and hexagonal lattices are displayed in left and right columns respectively. The upper panels [(a)(b)(e)(f)] are for a small filling ratio and the lower panels [(c)(d)(g)(h)] are for a high filling ratio. The numerically calculated $\partial \varepsilon_{\mathrm{eff}} / \partial u_{\mathrm{xx}}$ and $\partial \varepsilon_{\mathrm{eff}} / \partial u_{\mathrm{xy}}$ are calculated using the finite difference approach and are marked as black open circles. The analytic results (red lines) are calculated using Eq. (5) for the square lattice and Eq. (6) for the hexagonal lattice. The blue lines show analytic results according to Eq. (7), which carries correction terms for higher filling ratios. We note that high filling ratio correction terms are not required for $\partial \varepsilon_{\mathrm{eff}} / \partial u_{\mathrm{xy}}$.

FIG. 4. (color online) This figure compares the force density at site $(i,j)$ [ $\vec{f}(i,j)$ ] of the square lattice with $r_c = 0.3a$ for the $E_z$ polarization calculated using different approaches. The upper row is for the x-component and the lower row is for the y-component. The left panels are the results of the microscopic model, meaning that the fields calculated numerically for the system shown in Fig. 1(a) and the lattice structure are explicitly considered; and these results are by definition the correct results. Using the fields in the effective medium, the middle panels show force density results given by the Maxwell tensor, and right panels show the counterparts according to the Helmholtz tensor. In this polarization, the Maxwell and Helmholtz tensors are equal coincidentally and they give the same results as those calculated with the microscopic structure taken explicitly into account.

FIG. 5. (color online) The corresponding results (as those in Fig. 4) for the $H_z$ polarization. For this polarization, the Helmholtz tensor results are different from the Maxwell tensor results if we use the effective-medium calculated fields. Only the Helmholtz tensor which contains electrostrictive term gives the correct $\vec{f}(i,j)$ profile as shown in the left panel. The results show that based on the fields of effective medium, we must use the Helmholtz tensor in order to compute the force density correctly inside a medium.

FIG. 6. (color online) This figure compares the x-component of the force density $\vec{f}(i,j) \bullet \hat{x}$ for the square lattice with $r_c = 0.3a$ (left column) and the hexagonal lattice with $r_c = 0.279a$ (right column) calculated using different approaches for the $H_z$ polarization. We note that these different $r_c$ for the square and hexagonal lattice give the same $\varepsilon_{\text{eff}} = 1.57$ . The systems have slab geometry and the insets in panels (a) and (c) show pictorially the microscopic structure and the incidence angle of the plane wave is set to be 30 degree. The correct results (calculated using the microscopic lattice model) are shown as open squares. The force density has a jump at the boundary as shown in panels (a) and (c) and the jump is described well by the Helmholtz tensor results (shown as blue triangles) and the Maxwell tensor fails. The lower panels show that in the interior, the Helmholtz tensor also gives the correct results but not the Maxwell tensor. By comparing Fig. 6(b) and Fig. 6(d), we also note that the force density inside the slab is different for the square and the hexagonal lattice, even though $\varepsilon_{\text{eff}}$ is the same.

FIG. 7. (color online) The same results as shown in Fig. 6, but for a high filling ratio with $\varepsilon_{\text{eff}} = 3.08$. Again, the Helmholtz tensor gives better results. But in this case, we note that even the Helmholtz tensor does not give exactly the same result as that of the microscopic model, for reasons explained in the text.

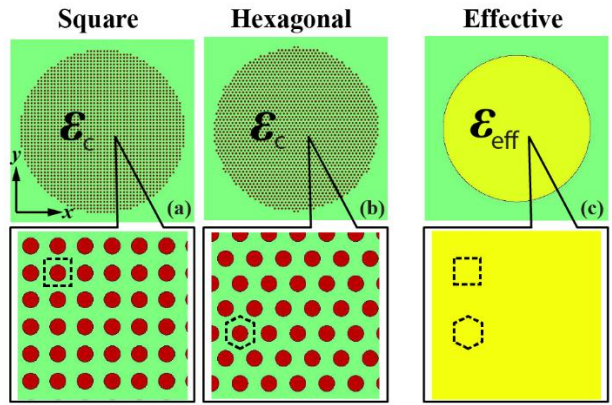

**Figure 1**

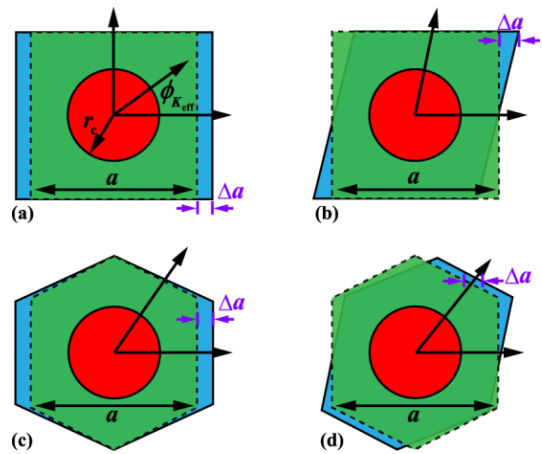

**Figure 2**

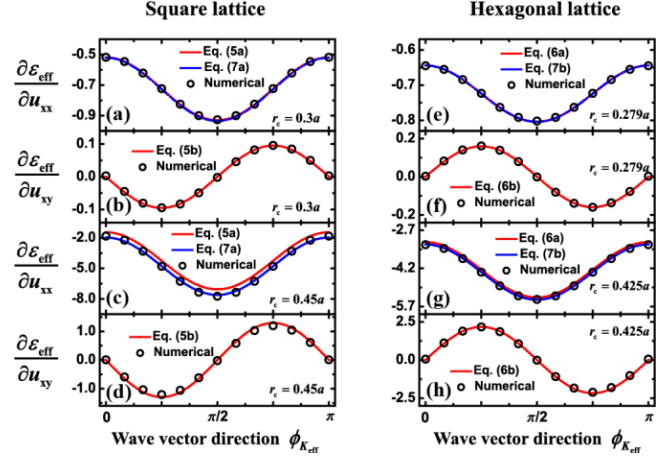

**Figure 3**

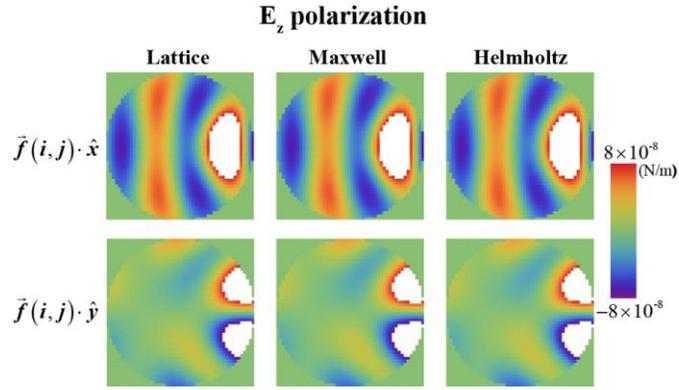

**Figure 4**

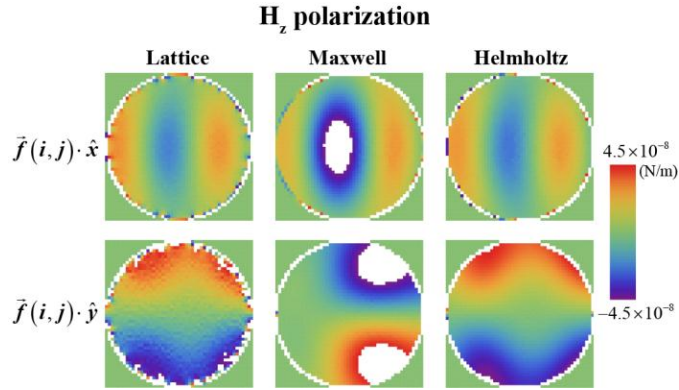

**Figure 5**

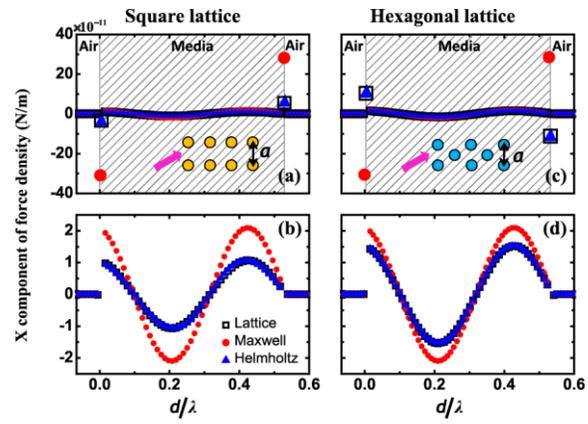

**Figure 6**

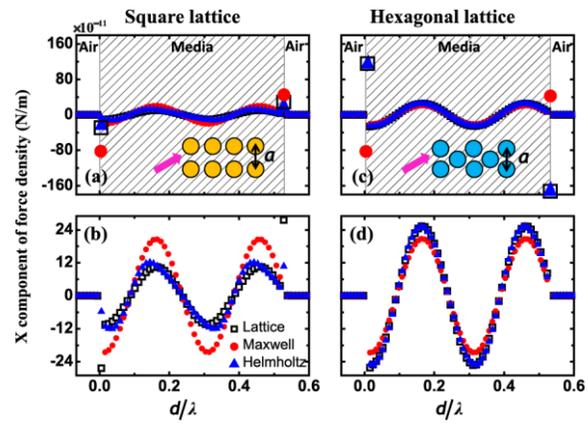

**Figure 7**